# Improving Smart Conference Participation through Socially-Aware Recommendation

Nana Yaw Asabere, Feng Xia, *Senior Member*, *IEEE*, Wei Wang, Joel J.P.C. Rodrigues, *Senior Member*, *IEEE*, Filippo Basso, and Jianhua Ma

***Abstract*—This research addresses recommending presentation sessions at smart conferences to participants. We propose a venue recommendation algorithm, Socially-Aware Recommendation of Venues and Environments (SARVE). SARVE computes correlation and social characteristic information of conference participants. In order to model a recommendation process using distributed community detection, SARVE further integrates the current context of both the smart conference community and participants. SARVE recommends presentation sessions that may be of high interest to each participant. We evaluate SARVE using a real world dataset. In our experiments, we compare SARVE to two related state-of-the-art methods, namely: Context-Aware Mobile Recommendation Services (CAMRS) and Conference Navigator (Recommender) Model. Our experimental results show that in terms of the utilized evaluation metrics: precision, recall, and f-measure, SARVE achieves more reliable and favorable social (relations and context) recommendation results.**

## I. Introduction

ATTENDEES at conferences are likely to have diverse research interests within broad research disciplines [1]. Academic conferences and workshops do not just serve as platforms to present research, but also aim to connect researchers/participants in the same domain and foster potential collaborations. The schedule of multiple and parallel tracks at academic conferences makes it difficult to identify which sessions may include participants with similar research interests. Additionally, the schedule may change due to the non-attendance by presenters. Thus participants may end up moving between session rooms.

One goal for event participation is to achieve high social capital and effective social learning. Social capital can be interpreted as a function of ties between actors in a social network [2]. Specifically, social capital can involve academic collaboration networks, where the actors are researchers, the friendships are collaborations, the events are conferences, the organizers are program committee members and the participants are authors [2].

Information extraction involves the integration of data from different sources. Such techniques lead to the generation of communities through the adaptation of data mining, information retrieval and recommendation techniques, which enable users to identify potential contacts for report sharing and community organization [3]. Recommender systems collect information concerning the preferences of users for a set of items. They use different sources of information and provide users with predictions and recommendations of items [4]. Mobile multimedia recommender systems [5] incorporating context and social awareness could support generating presentation sessions for participants.

This research addresses recommending presentation sessions at smart conferences to participants with the goal to enable the achievement of high social capital and successful social learning at conferences. We posit that participation in conferences can be improved through the integration of mobile technological devices, recommender system techniques, contextual information and social properties to enhance social awareness at such events.

The integration of Collaborative Filtering (CF) [6], Content-Based Filtering (CBF) [7] and Hybrid (H) [8] recommender systems which integrate users and items for generating recommendations can incorporate context [9]-[11] and mobile social networking properties [12] (see Table I). These advancements enhance the generation of reliable, trustworthy and efficient recommendations for users. Our proposed algorithm, *Socially-Aware Recommendation of Venues and Environments* (*SARVE*), recommends conference presentation session venues and environments to participants by utilizing socially-aware, and distributed community detection techniques. *SARVE* aims to detect and recommend conference

This work was partially supported by the Natural Science Foundation of China under Grant No. 60903153 and No. 61203165, Liaoning Provincial Natural Science Foundation of China under Grant No. 201202032, the Fundamental Research Funds for the Central Universities, the Instituto de Telecomunicações, Next Generation Networks and Applications Group (NetGNA), Portugal, and National Funding from the FCT – Fundação para a Ciência e Tecnologia through the Pest-OE/EEI/LA0008/2013 project.

Nana Yaw Asabere, Feng Xia, and Wei Wang are with School of Software, Dalian University of Technology, Dalian 116620, China.
Joel J.P.C. Rodrigues and Filippo Basso are with Instituto de Telecomunicações, University of Beira Interior, Portugal.
Jianhua Ma is with Faculty of Computer and Information Sciences, Hosei University, Japan.
Corresponding author: Feng Xia; E-mail: f.xia@ieee.org

presentation session venues that are important and related to the research interests of participants.

TABLE I

CATEGORIES OF TRADITIONAL RECOMMENDER SYSTEMS

| Recommender System | Brief Description |
|---|---|
| CF | The CF approach gathers ratings on the items by a large number of users and makes recommendations based on the interest patterns of other users. The CF approach is based on the assumption that a user would usually be interested in those items preferred by other users with similar interests. |
| CBF | The CBF approach examines the content information related with the items and users in order to denote users and items using a set of features. To recommend new items to a user, CBFs match their representations to those items known to be of interest to the user. |
| H | H combines the CF and CBF as well as other recommender algorithms/systems to reduce challenges and problems such as cold-start and data sparsity. |

*SARVE* obtains information concerning research interests, physical contact durations and contact frequencies of individual conference participants in order to determine their preference similarities and social tie strengths in terms of research. To detect different communities consisting of presentation sessions at the conference, *SARVE* considers different sources of information including: (i) context (locations and times of different presentation sessions and available times and locations of participants), (ii) personal (research interests of participants) and (iii) social (tie strengths between the presenters and the other participants as well as degree centrality of the presenter). The distributed community detection algorithm organizes and allocates the participants into different and common communities/sessions at the conference.

Our contributions in this work include the following:

- By exploiting correlation, social ties pertaining to presenters and participants, and the degree centrality of presenters, we develop methods and procedures to detect different presentation sessions (communities) to attend.
- We also determine the extent of relationships between attendees and presenters and the popularity level of the presenters.
- Our method quantifies the extent of research tie (weak or strong) relationships among presenters and participants, and the popularity level of presenters to generate social relation recommendations.
- We propose a distributed community detection algorithm that recommends presentation session venues to participants based on high research interest similarity, strong social relations and the matching of contextual information between the presenters and participants at the conference venue.
- We compare the approach with existing state-of-the-art methods.

The rest of the paper is organized as follows. Section II reviews related work on social relation recommendations, social context recommendations and conference session recommendations. The operational concept and algorithmic design of our *SARVE* are discussed in Sections III and IV respectively. In Section V, we present our experimental evaluations. Finally, Section VI concludes the paper.

## II. RELATED WORK

There are recommender systems that do not account for contextual information [9]. Next, we discuss recommender systems and algorithms involving the utilization of contextual, social information and social relationships.

### A. Social Relation Recommendations

Social recommendation methods, which consider only one kind of relationship in social networks, face data sparsity (users rating a small proportion of items out of a larger number of available items) and cold-start problems (new user and new item problems) [4], [13], [14]. To address this issue, Chen *et al.* [15] proposed a recommendation method based on multi-relational analysis. They combined different relation networks by applying optimal linear regression analysis, and then, based on the optimal network combination, they put forward a recommender algorithm combined with a multi-relational social network.

Guy *et al.* [16] studied personalized recommendation of social software items, including bookmarked web-pages, blog entries and communities. They focused on recommendations that are derived from the user's social network. They compared recommendations that are based on the user's familiarity network and his/her similarity network. Based on a survey involving 290 participants and a field study including 90 users, the authors found out that familiarity network in terms of relationships is an innovative basis for social recommendation.

Zhou *et al.* [17] facilitated knowledge and sharing enhanced collaborative learning by considering two important factors, namely: user behavior patterns and user correlations. Within a task-oriented learning process, they described relations of learning tasks, activities, sub-tasks and tasks in communities. Based on these relations as well as relevant algorithms, they developed an integrated mechanism to utilize both user behavior patterns and correlations for the recommendation of individual learning actions.

Chen *et al.* [18] proposed a method by using clustering, *SimRan*k and adapted *SimRank* algorithms to recommend matching online dating candidates. *SimRan*k scores the similarities of users based on how similar the people they have contacted are, in terms of social network connections. The adapted *SimRank* scores the similarity of users based on similarity between their contacts in the cluster. The authors found out that social (relations) information improve recommendations for online dating networks. They also found out that their social recommendation results could be improved through the combination of user profiles and preferences.

### B. Social Context Recommendations

Existing social recommendation approaches consider social network structures, but social context has not been fully considered [13], [19]. Due to the social characteristics/features of users, it has become necessary as well as challenging to fuse social contextual factors into social recommendation procedures [4], [13]. Jiang *et al.* [19] identified that individual preference and interpersonal influence are important factors for social recommendations.

Ma *et al.* [13] proposed a factor analysis approach based on probabilistic matrix factorization to alleviate the data sparsity and poor prediction accuracy problems by incorporating social contextual information such as social networks and social tagging. Their approach performed better than other state-of-the-art methods, especially in circumstances where users had made few ratings.

Biancalana *et al.* [20] described a social recommender system that is able to identify user preferences and information needs, thus suggesting personalized recommendations related to Point of Interests (POIs) in the surroundings of a user's current location.

Liu *et al.* [21] investigated context aware movie recommendation tasks: (i) how to combine multiple heterogeneous forms of user feedback? (ii) how to cope with dynamic user and item characteristics? and (iii) how to capture and utilize social connections among users? They proposed to use ranking techniques based on matrix factorization models.

### *C. Conference Session Recommendations*

In order to suggest context-aware and personalized information, intelligent processing techniques are necessary [22]. Determining user interest can enable the suggestion of contextualized and personalized information [9], [22], [23]. Characterizing social interaction features and contextualized social relations of users can support social activity organization [24].

In terms of recommendations of presentation/talk session venues at conferences, Pham *et al.* [1] presented the Context-Aware Mobile Recommendation Services (*CAMRS*). They augmented the current context with the academic community context of participants which was inferred by using social network analysis and link prediction on a large-scale of co-authorship and citation networks of participants. By combining the dynamic and social context of participants, they were able to recommend talks and people (presenters) that may be of interest to a particular participant.

Farzan and Brusilovsky [25] also presented a social information access system that helps researchers plan talks they wish to attend at large academic conferences. They attempted to address the problem of collecting reliable feedback from the community of conference participants. Following a "do it for yourself" approach, their system encourages participants to add interesting talks pertaining to their individual schedules and uses scheduling information for social navigation support.

Hornick and Tamayo [26] introduced a Recommendation Engine called *RECONDITUS* which conjoins decomposition of items and users. *RECONDITUS* recommends items from a new disjoint set to users. It requires no item ratings, but operates on observed user behavior such as past conference session attendance.

Our previous work [27] proposed a socially-aware venue recommendation algorithm which fuses location and time contextual data. The approach was evaluated using precision and recall metrics. The method outperformed other state-of-the-art methods. In this paper, we further evaluate *SARVE* with an additional metric (f-measure) to address reduction in data sparsity and cold-start problems. This work also addresses distributed community detection techniques and degree centrality with relevant diagrams. To the best of our knowledge, the generation of social recommendations for conference participants using a combination of Pearson correlation, social ties, contextual information and degree centrality (popularity level) of a presenter is only accomplished in [27].

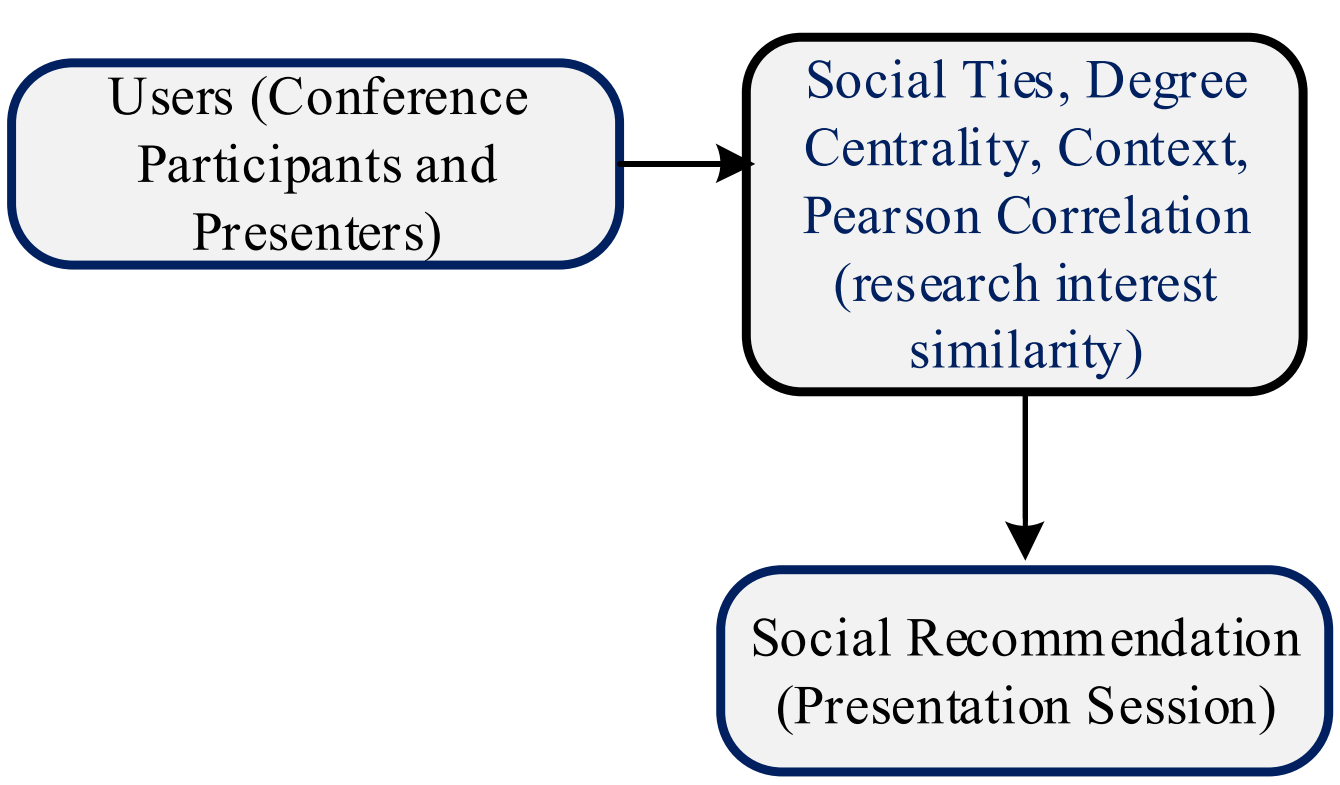

Fig. 1. Basic recommendation procedure flow of SARVE.

## III. SARVE Operational Concept

The premise of this work is that the incorporation of social properties in addition to context and traditional recommender system techniques will be more beneficial in terms of enhancing the generation of effective social recommendations for conference participants and the reduction of data sparsity and cold-start problems [4], [13], [14]. This is because the social properties of nodes/users in a network are important features to consider when analyzing social data for an effective output such as socially-aware recommendation. Fig. 1 shows the basic recommendation procedure of *SARVE* and thus depicts our motivation and innovation through the recommendation entities we utilize.

Fig. 2 shows that through the augmentation of relevant context, the *SARVE* algorithm generates both social relations and social context recommendations by respectively computing social tie, degree centrality and Pearson correlation. The similarity between an active presenter ($C_p$) and another participant ($C_x$) is measured as the tendency to rate tags (keywords) closely or similarly [28]-[30].

The interests of mobile device users (conference participants) can change at any time because of the changes in their surrounding environments, e.g. physical conditions, location, time and their community (smart conference). Therefore the recommendation service in *SARVE* considers static and dynamic user profiles.

Referring to Fig. 2, the upper-left side depicts an interactive scenario between the conference participants ($C_1$, …., $C_n$), who are the users and a $C_p$ at the smart conference. During the main conference (before presentation sessions begin), if a participant makes a social recommendation request to attend a relevant presentation session(s), *SARVE* utilizes relevant information to compute the Pearson correlation and social ties of the participant and all the presenters to ascertain high levels of research similarity and tie strength between them. Furthermore, *SARVE* computes the degree centrality of presenters to determine their popularity status/level at the smart conference and further integrates explicit contextual information of the participant, presenters and community, in order to accordingly generate social venue recommendations.

The *Conference Participant Collector* gathers and sends the tagged ratings of the individual conference participants to the *Conference Profile Engine* for the computation of user context. The *Social Tie Analyzer* computes the contact durations and contact frequencies between the $C_p$ and other participants to determine their tie strengths. The *Degree Centralizer* computes the social popularity of a $C_p$ with the other participants by measuring the extent of their direct connections and ties.

The *Contextual Post-Filtering* technique involves contextualizing recommendation outputs for participants based on their tagged ratings through traditional 2D procedures [9]. *SARVE* verifies and contextualizes the resultant location, time, user and social relations contexts of the smart conference community and participants. The post-filtering contextualization procedure, which involves context of users and the conference community, enables *SARVE* to generate social context and relation recommendations.

## IV. SARVE Design

This section includes our approach for computing similar research interests of presenters and participants using Pearson correlation coefficient. Then, we describe the methods of computing the social ties of presenters and participants and degree centrality of the presenters. We describe how we sense contextual information and match contextual relationships in *SARVE*.

### A. User Interests and k Most Similarity

By using their mobile devices, conference participants specify their research interests via specific keyword. In the implementation, a tag is a relevant keyword assigned to one or more research interests of a conference participant. Participants also enter the contact durations and frequencies between presenters and themselves.

CF algorithms are divided into memory-based and model-based approaches. Since our method employs a user-item database, the memory-based CF is more appropriate in comparison to model-based (which involves the design and development of a model such as machine learning for making intelligent predictions). CF uses two main methods: User-Based CF and Item-Based CF. In *SARVE*, we utilize User-Based CF because of the importance of the similarity of research interests among participants (users), rather than similarity of items. User-Based CF involves the following steps: (i) Look for users (presenters) who share the same tagged patterns with the active user and (ii) Use the ratings from those similar interest to calculate a recommendation for the active participant.

We utilize the Pearson correlation coefficient to identify and compute the $k$ most similarity between two users' (nearest neighbors) involving a presenter, $C_p$ and a participant, $C_x$. Each user is treated as a vector in the $m$-dimensional item space and the similarities between $C_p$ and $C_x$ are computed within the vectors.

After the $k$ most similar users have been identified through a user-item matrix, the User-Based CF technique generates a top-$N$ recommendation list for $C_x$ based on tagged rating similarities with $C_p$. Using (1), we compute the Pearson correlation between presenters and participants i.e. $C_p$ and $C_x$. In (1), $C_p$ and $C_x$ are represented as $c$ and $d$ respectively. Therefore, the similarity between $C_p$ and $C_x$ is denoted by $Sim(c, d)$. The tagged ratings of $c$ and $d$ for item $i$, (where $i \in I$ and $I$ is the set of items) are denoted by $r_{c,i}$ and $r_{d,i}$ respectively. The average ratings of $c$ and $d$ are denoted by $\bar{r}_c$ and $\bar{r}_d$ respectively.

$$Sim(c,d) = \frac{\sum_{i \in I}(r_{c,i} - \bar{r}_c)(r_{d,i} - \bar{r}_d)}{\sqrt{\sum_{i \in I}(r_{c,i} - \bar{r}_c)^2}\sqrt{\sum_{i \in I}(r_{d,i} - \bar{r}_d)^2}} \qquad (1)$$

Using (2), we set a threshold, $\gamma$ (to be determined in our experiment) for (1), to define the preference similarity between $C_p$ and $C_x$ in terms of tagged (keyword) ratings (1-5).

$$Sim(C_p, C_x) \geq \gamma \qquad (2)$$

The similarity values between $C_p$ and other participants have to fall within the defined threshold before such participants can be detected as members of the community where the presenter will be delivering his/her presentation.

### B. Tie Strength

The social relations among individuals are usually called social ties. Ties typically represent the existence of a substantial relationship between two individuals, for example, acquaintance and research familiarities [12]. Using (3), we measure and estimate the tie strength between a $C_p$ and $C_x$. In (3), $d_{Cp,\,Cx}(t)$ is the contact duration between the $C_p$ and $C_x$ in the time frame $T$ and $\lambda_{Cp,\,Cx}$ is their contact frequency (i.e. the number of times $C_p$ and $C_x$ have been in contact within the time frame $T$).

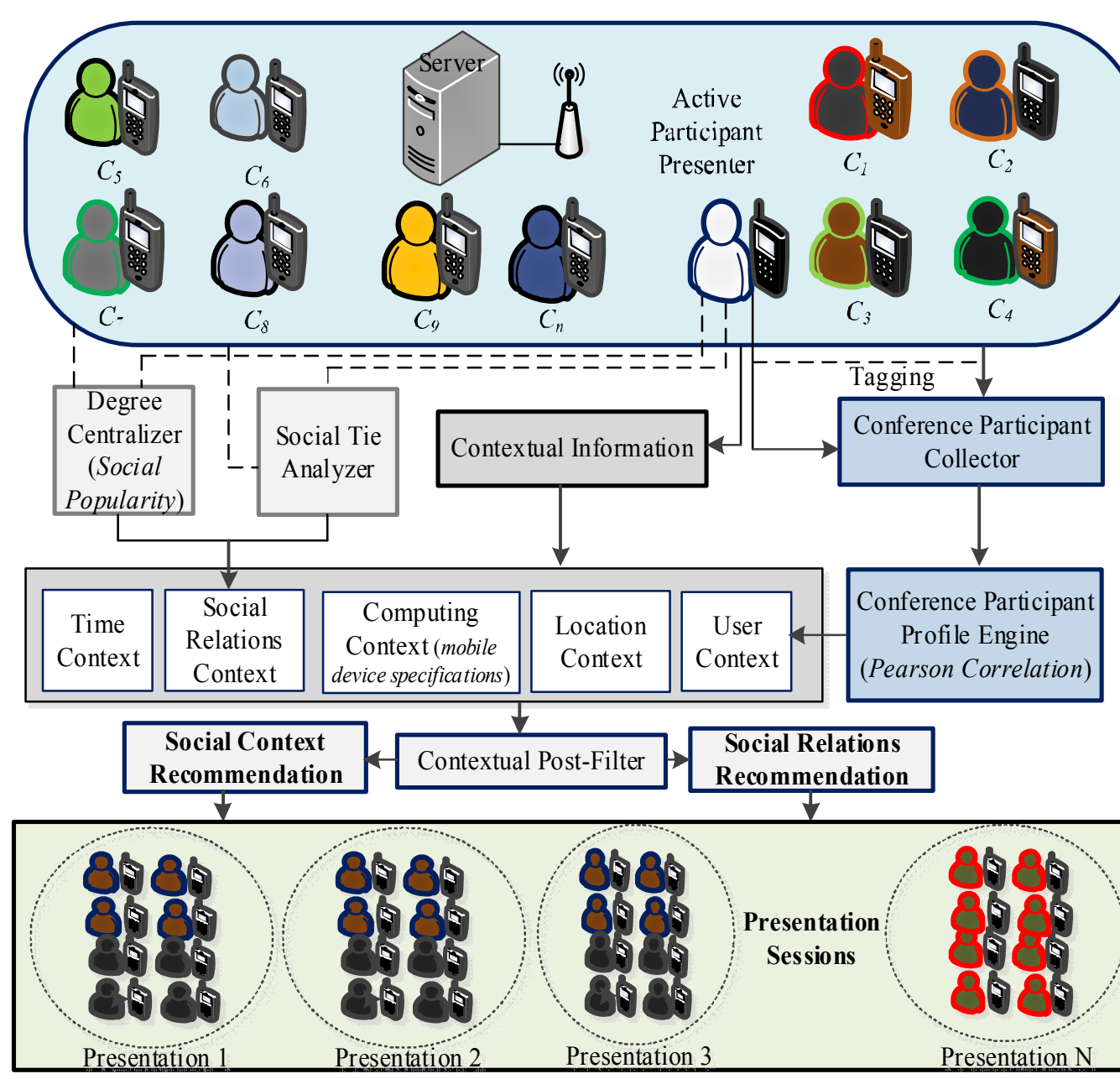

Fig. 2. SARVE recommendation model.

$$SocTie_{C_p,C_x}(t) = \frac{(\lambda_{C_p,C_x} \times d_{C_p,C_x(t)})}{T} \qquad (3)$$

Using (4), we set a threshold, $\beta$ (determined empirically) for (3) to define the tie strength between $C_p$ and $C_x$. The social tie values between $C_p$ and other participants have to fall within the defined threshold before such participants can be detected as members of the community where the presenter will be delivering his/her presentation. For example, if a participant specifies that his contact frequency at the conference with a $C_p$ is 5 ($\lambda c_p,c_x = 5$) in a duration of 60 minutes ($dc_p,c_x(t) = 60$) and conference time frame of 660 minutes ($T = 660$), then by using (3), their social tie result will be computed as $SocTiec_p,c_x(t) = (60 \times 5)/660 = 0.45$. Such a social tie result may be low or high in accordance to a particular threshold.

$$SocTie_{C_p,C_x}(t) \geq \beta \qquad (4)$$

### C. Degree Centrality

Degree Centrality measures the number of direct connections and ties that are associated with a given user/node. A user associated with more social ties represents a more important location for a community in a network than a user with fewer or no social ties. A user with high degree centrality maintains contact durations and frequencies with other users in the network. Such users can be seen as the most active and popular with a large number of links in comparison to other users in the same network [12], [31]. In *SARVE*, we assume that a $C_p$ that has a higher number of social ties and connections with other participants are popular and consequently their popularity can be used as added incentives to generate effective presentation session recommendations for the participants.

$C_p s$ that maintain few or no social ties and connections are described as unpopular within the network. The degree centrality for a given $C_p$, includes a function $a$, where $a(C_p,C_x)$ = 1, if a direct link exists between $C_p$ and $C_x$. Degree centrality for a given $C_p$ is therefore computed as [31]:

$$C_D(C_p) = \sum_{1}^{N} a(C_p, C_x) \qquad (5)$$

where $N$ is the total number of users/nodes in the network. Fig. 3 illustrates an example of the degree centrality of a presenter at the smart conference community. User (presenter) 4 has the most direct connections amongst the other users/nodes and hence has the highest degree centrality.

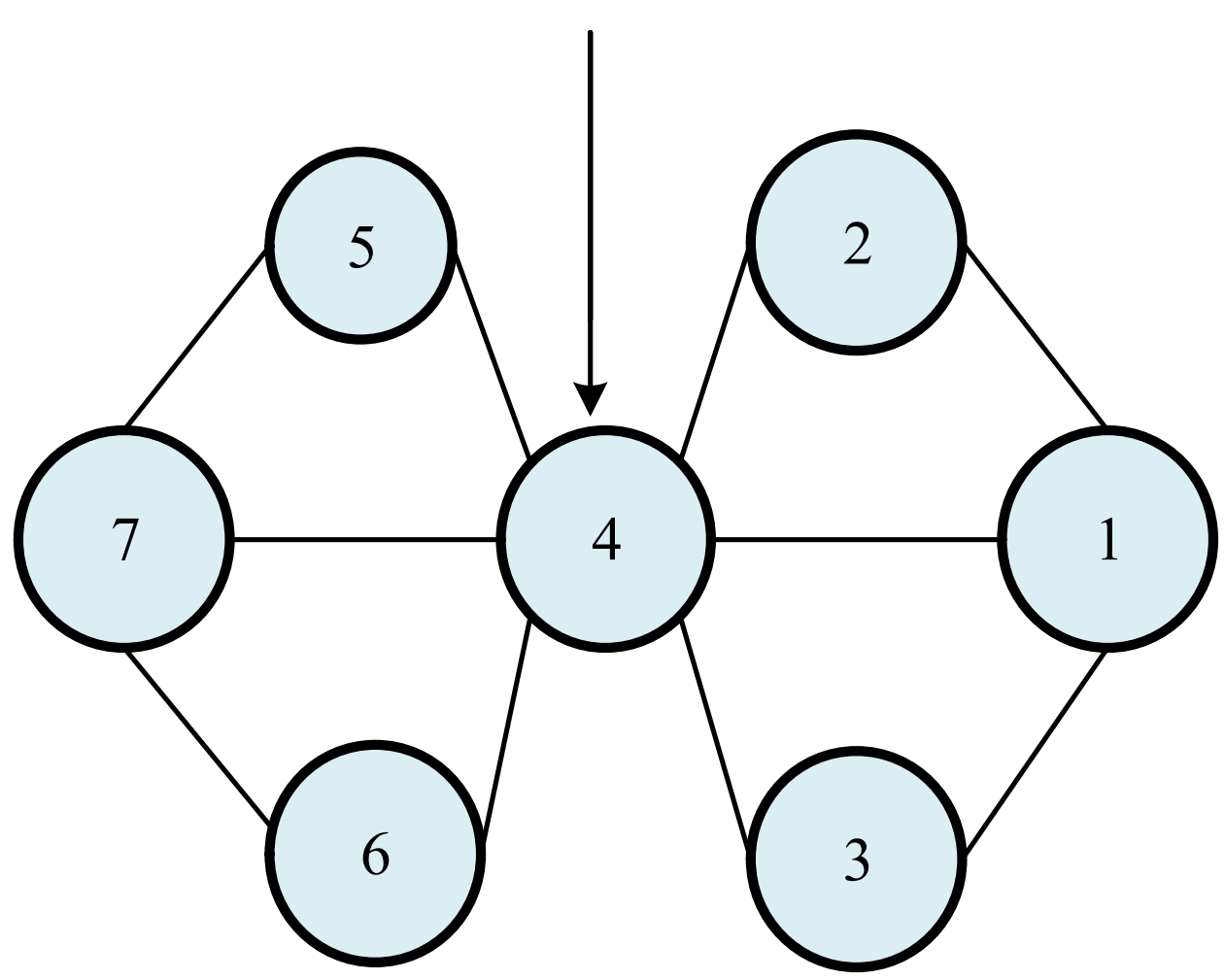

Fig. 3. An example of the degree centrality of a presenter in a smart conference where presenter 4 is most popular.

### D. Contextual Information Sensing

A specific definition and model of context in recommender systems can expedite what constitutes context and can facilitate the usage of contextual data across various applications. Context is often defined as an aggregate of various categories that describe the setting in which recommender systems are deployed. *SARVE* utilizes five types of contexts, namely: computing, location, time, user and social relations.

*Computing Context*: *SARVE* requires standard android smartphone specifications. Information pertaining to these specifications is sensed implicitly. Specifically, through a request header, *SARVE* captures the specifications about the smartphones belonging to participants to ascertain whether a social recommendation is possible through their devices.

*Location Context*: Global Positioning System (GPS) and Wi-Fi are available in modern mobile phones. These technologies enable tracking human location behavior at scales that were previously unattainable [23], [32], [33]. When the exact location is required, users can manually input their location type [34] or participants can label places visited with departure times [35]. As *SARVE* involves the detection of exact venues of presentation sessions, we utilize an explicit procedure to sense the precise locations of presenters and participants at the smart conference.

*Time Context*: Time context usually involves the exact date and time information. Time can either be precise (e.g. within five minutes) or vague (e.g. within a week, sometime in a month or in the coming semester/academic year). Time and other contexts can be combined [23]. For example, Rosa *et al.* [36] combined several multimedia sources in a mobile recommender system for events. Their approach was based on few weighted context-aware data-fusion algorithms. They presented a demonstrative deployment procedure which utilized context-aware data such as location, time, user sharing statistics and user habits.

Timestamp data can be is captured from available data such as a learning schedule. For instance, in [37], the Context-aware Adaptive Learning Schedule (*CALS*) provides a learning schedule that allows users to enter their time data, for planning their leaning activities. Similar to *CALS*, *SARVE* also provides a smart conference schedule with available presentation session dates and times to enable participants to enter their specific time data for available presentation sessions.

*User Context*: We sense the context of the users (presenters and participants) through explicit tagging of their research interests. We compute the research interest similarities between the presenters and participants using correlation.

*Social Relations Context*: We sense the social relations context of the presenters and participants through the computation of their social ties and degree centrality of presenters. These computations determine their ties strength and allow various participants to join a presentation session based on his/her tie strength with a presenter(s) as well as popularity level of the presenter(s).

### *E. Contextual Relationships Matching*

In social tagging systems, a user's tagging and commenting activities generate relations involving more than two types of entities [38] and the posts (that is, each tag produced by a user for an item) are classified as third order data [39], [40]. Yin *et al.* [38] highlighted that this classification is further considered as a triple (user-tag-item) as shown in Fig. 4.

We adopt the model called the *Bipartite graph between relations and entity types* in [38] and use it to establish social relationships between $C_p$ and $C_x$ in terms of context. This facilitates the generation of social recommendations based on the *k* most similarity and social tie results of participants obtained from (1) and (3) and subsequent computed threshold values from (2) and (4). An example of four relations on five entity types in a social tagging system is depicted in Fig. 4.

In Fig. 4, A1 is the social network context (user-user), A2 is the comment context (user-comment-item), A3 is the item-content context (item-content feature) and A4 is the tag post context (user-tag-item). If the results of (1) and (3) depict that $C_p$ and $C_x$ have *k* most similarities and strong social ties, then the presentation (*Item* ($P_n$)) annotated with a tag by $C_p$, based on a comment feature about the location and time of the presentation and content feature will be the identified and detected presentation community for $C_x$. It must be noted that the extent of social relationship in terms of context between $C_p$ and $C_x$ can only be generated based on the results of (1) and (3) i.e. if the research interest similarities and social ties of $C_p$ and $C_x$ doesn't fall within the computed threshold results, a social relationship cannot be established using Fig. 4.

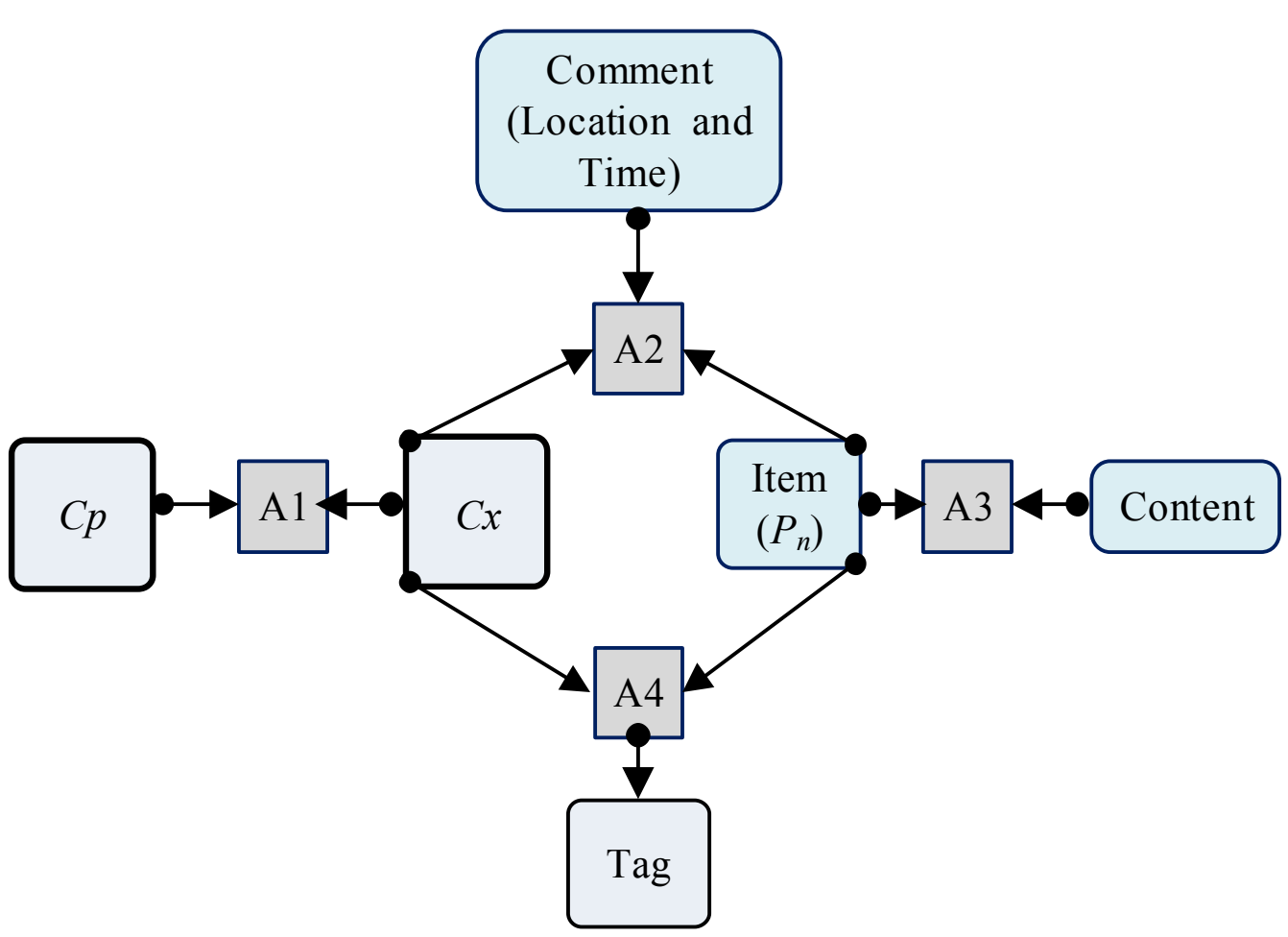

Fig. 4. Bipartite graph between conference participant relations and entity types.

### *F. Community Detection*

There are two types of methods used to detect a community in Mobile Social Networks (MSNs): *centralized* and *distributed* community detection techniques. In the *centralized* technique, full knowledge of the whole MSN and its ties are needed, while in the *distributed* technique, each node or user is able to detect the community it belongs to [12].

The majority of community detection algorithms require global information or centralized control. Centralized community detection algorithms scale very poorly in terms of the number of nodes and edges present in the MSN. Such algorithms are infeasible in large-scale real networks due to computation and accessibility [41], [42].

Collingsworth and Menezes [41] proposed a Self-Organized Community Identification Algorithm (SOCIAL), based on local calculations of node entropy and enables individual nodes to independently decide the community they belong to. Chen [42] proposed a distributed algorithm based on information diffusion. In [42], it was revealed that information in the human society can allow people to understand the emergence of a community structure. Huang *et al.* [43] also proposed a distributed community detection algorithm in which communities are detected for mobile learners based on their learning networks and research interests.

We propose a distributed community detection algorithm in which users (participants) independently detect related presentation session venues (communities) through the generation of social context recommendations and social relation recommendations. We detect distributed communities of participants based on their research interests, social ties and tagged ratings as well as the social popularity of presenters.

In Fig. 5, presentation sessions are detected for a target user (participant) represented as $C_x$. Fig. 5 shows that a target user is allocated a presentation session based on his/her community preference resulting from strong ties and high research similarity level with other participants (who are presenters) at the smart conference. The presenters are part of the communities in which $C_x$ is attached to, therefore $C_x$ is recommended a presentation session facilitated by a $C_p$ as depicted on the right side of Fig. 5.

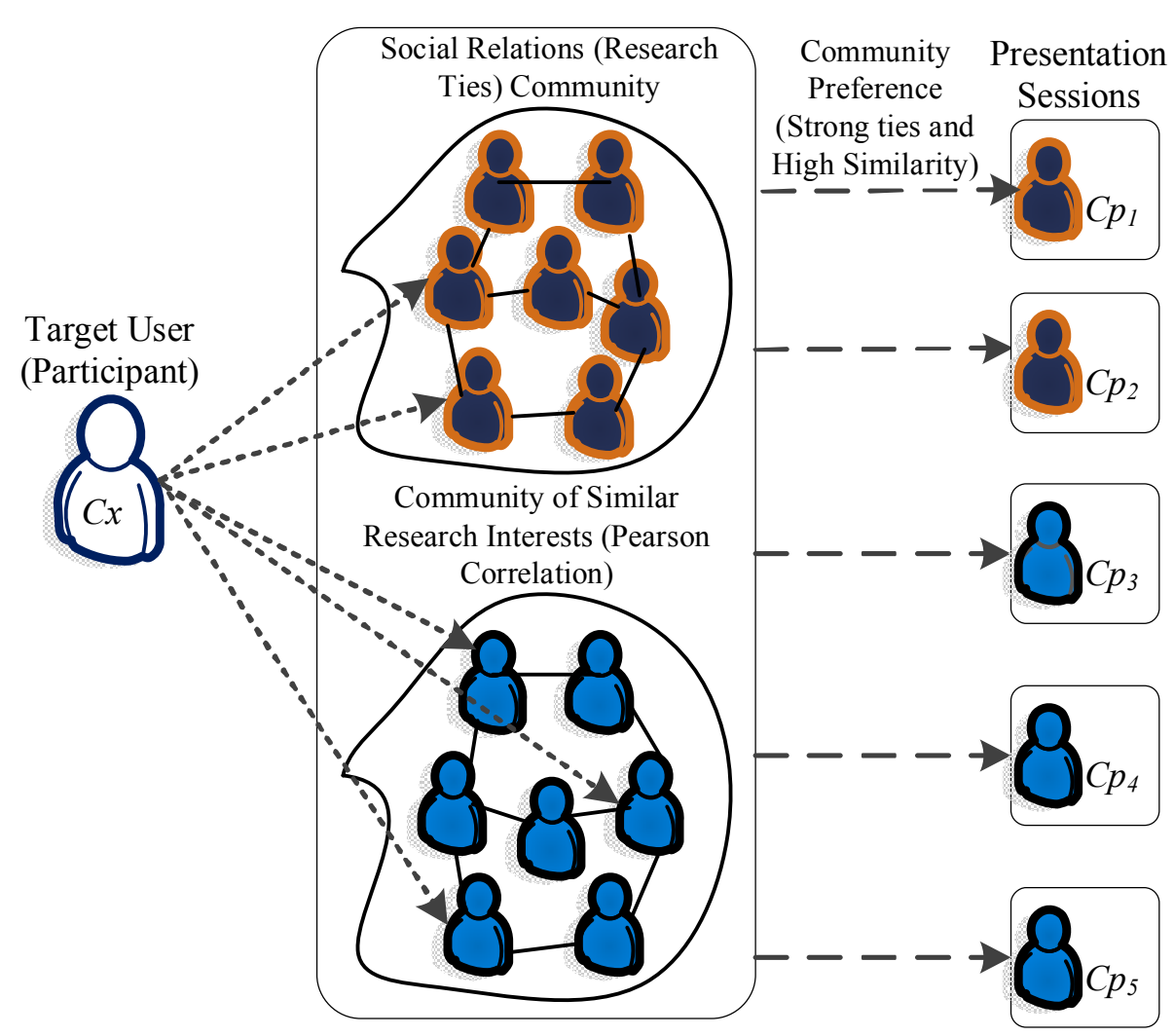

Fig. 5. Presentation session (community) detection for conference participants.

Our distributed community detection algorithm declares and initializes integer, floating and string variables. The integer variables consist of *i, j, m, n* and *z*, where *i* and *j* are initialized to a value of 0 and used for comparison of transactions in the array of presenters of size [*m*] and participants of size [*n*] both consisting of tagged ratings and social information through *for* loops based on incremental transactions. These steps are depicted in 1-8. Steps 9 and 10 compute the Pearson correlations between the participants and presenters. Based on the results of the Pearson correlation computations, steps 11-17, compare the contextual parameters of participants and presenters and accordingly generates social context recommendations. The final steps (18-28) compute the social ties of the participants and presenters, as well as the degree centrality of presenters and accordingly generate social relation recommendations.

## V. Evaluation

This section presents the performance and evaluation of relevant benchmarking experiments. We introduce the dataset utilized and our experiment setup. Then, we present the evaluation metrics to test the performance of our algorithm. Finally, we present our experimental analysis and results.

### *A. Overview*

Both online and offline evaluations were conducted. Different features of recommender algorithms were considered in the evaluations [44]-[46]. An online evaluation is challenging, so a simulated online process where the system makes recommendations or predictions and the user uses the recommendations or corrects the predictions was used.

We simulated the 2012 International Conference on Web-Based Learning (ICWL 2012) which involved recording historical user data in order to obtain the knowledge of how a user (participant) would rate an item or which recommendations a user would act upon. The dataset included 60 presenters, each with five contacts and with individual contact durations and frequencies used for social tie and degree centrality computations. Additionally, the interests of the presenters were acquired through the keyword tags obtained from the titles of their presentations. Contextual information involving the location of presentations, time of presentations and date of presentations are also available in the dataset.

To identify research interests, social and contextual information of participants, we gathered data from 78 members/students of the School of Software, Dalian University of Technology, China. The members/students were instructed to select/annotate keywords of interest as well as social and contextual information (available time and present location) in relation to the simulated conference (ICWL 2012).

At ICWL 2012, presentations for full papers were 20 minutes plus 5 minutes for questions, for short papers, 15 minutes plus 5 minutes for questions and for workshop papers, 15 minutes plus 5 minutes for questions. There were two main Conference session venues for different presentations at multiple times in

**Algorithm**: Pseudocode for detecting and recommending presentation session venues

```
1: // Declare and Initialize Variables
2:  i, j, m, n, and z;                    // integer variables
3:  pearson_threshold_val, soctie_threshold_val,
    social_tie[z]  deg_cent_threshold and  Pearson[z];
                                          // floating variables
4:  location[n], time[n];        // string variables
5:  Participants [n];            // array of Participants of size n
6:  Presenters[m];               // array of Presenters of size m
7:  for (i=0 to i<n increment i)
8:      for (j=0 to j<m increment j)
9:          Compute Pearson correlations using Eq. (1) and
            store  in Pearson[z]
10:         if (Pearson[z] ≥  pearson_threshold_val) then
11:             Compare contextual parameters;
12:           if (Presenter[j].location == Participant[i].location)
                AND (Presenter[j].time == Participant[i].time)
                then
13:               // Generate Social Context Recommendation
14:               Assign Participant[i] to Presenter[j];
15:           end if
16:         end if
17:         increment z
18:         Compute Social Ties using  Eq. (3) and store in
            social_tie[z]
19:     Compute Degree Centrality of Presenters using  Eq. (5)
20:         if (SocTie_Cp, Cx(t) ≥  soctie_ threshold_val) OR
            (Participant[j].deg_cent≥deg_cent_threshold) then
21:             Compare contextual parameters;
22:          if (Presenter[j].location == Participant[i].location)
              AND  (Presenter[j].time  ==  Participant[i].time)
              then
23:             // Generate Social Relations Recommendation
24:             Assign Participant[i] to Presenter[j];
25:          end if
26:        end if
27:     end for
28: end for
```

the ICWL 2012 conference. These include Building 1 - George Enescu (GE) Hall (Room A) and Building 1 - Mircea Eliade (ME) Hall (Room B).

The dataset was divided in the training set and the test set. We allocated 80% of the data for training and 20% for testing [45]. The contact durations ranged from 5-80 minutes (Fig. 6a). Fig. 6(b) depicts the tagged ratings trends in the dataset. The ratings of participants ranged from 1-5 and the number of participants who annotated tags with specific ratings are also shown in Fig. 6(b). Furthermore, Fig. 6(c) shows the dataset information involving the contact frequency trends between participants and presenters with their frequencies (times of contact) ranging from 1-7.

We assumed a time frame (*T*) of 12 hours (720 minutes) for the total duration of the smart conference. Using (3), we computed $SocTie_{c_p,c_x}(t) = (80 \times 7)/720$ and obtained a result of 0.8 as the highest positive and effective recommendation based on strong social ties between presenters and participants. In

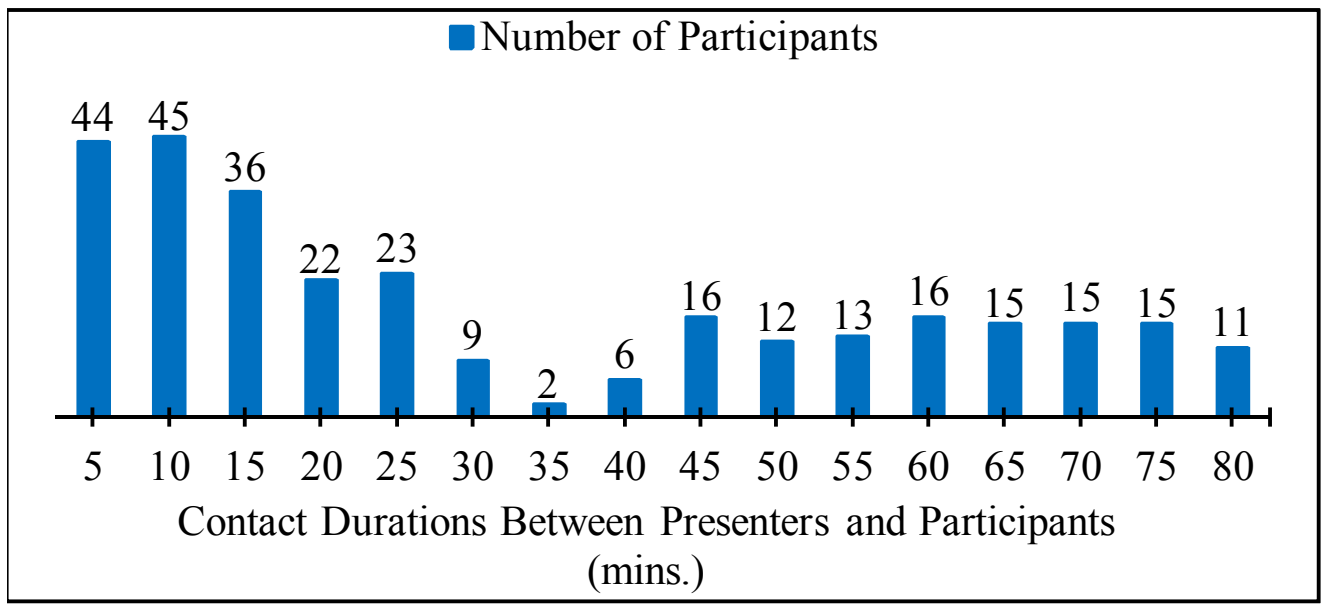

(a) Contact duration trends

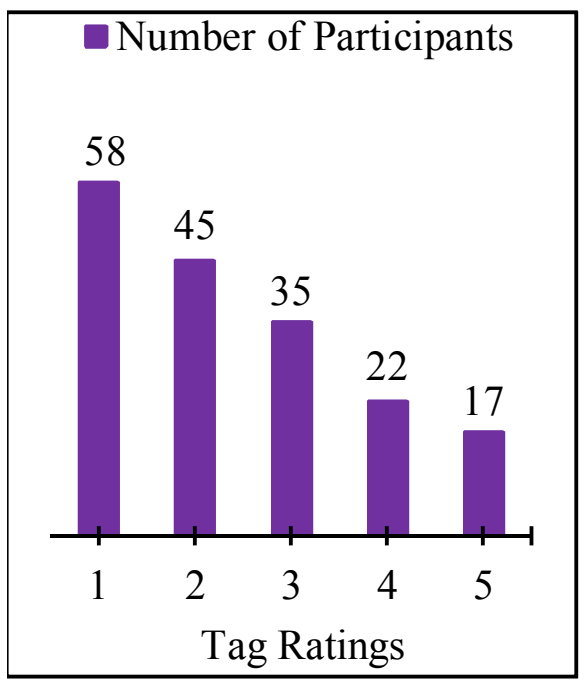

(b) Tagged rating trends

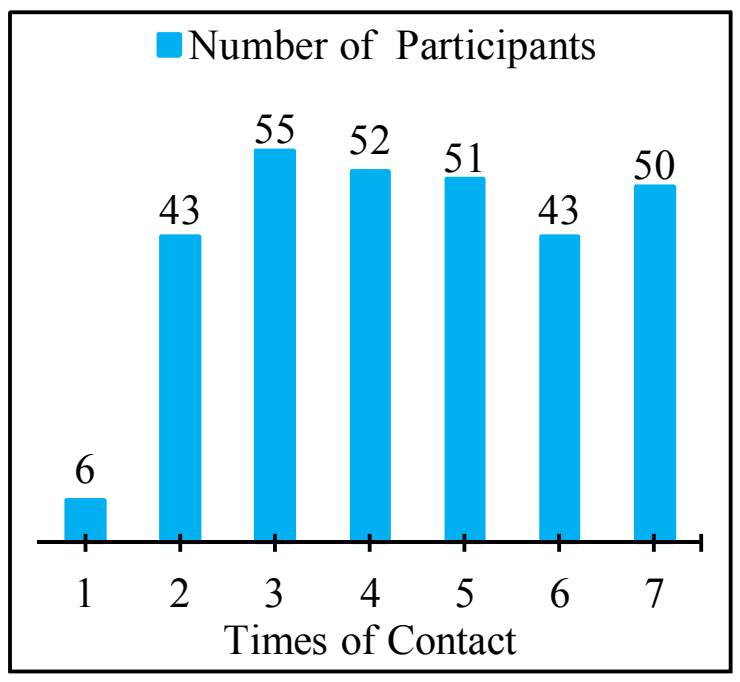

(c) Contact frequency trends

Fig. 6. ICWL 2012 dataset.

*SARVE*, the tie strength and tagged rating similarity levels between participants and presenters determine the quality of a social recommendation. Social ties between 0.5-0.8 and Pearson correlation between 0.6-1.0 in the dataset generated more effective social recommendations in terms of quality. Computed recommendation values that fell within these thresholds constituted the participant's priority recommendation list. Computed recommendation values below the thresholds were thus considered weak recommendations. We set the range for recommendation based on the social ties (relations) as $0 \leq SocTie_{c_p,c_x}(t) \leq 0.8$ and allocated a social relations recommendation threshold of 0.5 and above in accordance to the dataset.

## B. *Baseline Methods for Comparison*

We compared *SARVE* to the work in [1] and [25] where B1 and B2 denote methods of [1] and [25] respectively. B1 and B2 involved recommendations for conferences presentation sessions which are quite similar and related to *SARVE*, and paved the way for a methodological comparison. We briefly describe the baseline methods of [1] and [25] below.

The approach in [1] followed the Multidimensional Recommendation Model (*MRM*), where the preference data are decomposed according to time and location dimensions. By using link prediction methods on large-scale social networks, the community of users (social context) was identified and combined with dynamic preference data which was used in the recommendation service. The recommendation method in [1] was evaluated using relevant datasets such as simulation of a conference (ICWL 2010) and the utilization of the digital library DBLP.

The recommendation method in [25] explored the value of social navigation and social search technologies in the context of conference attendance planning. The Conference Navigator (Recommender) Model in [25] was designed to assist the conference attendees in making decision about which talks/presentations to attend. The approach in [25] employed the collective wisdom of the community based on feedback collected from a community of users with similar interests and social navigation support techniques. Activities were introduced to users (participants) who provide reliable indication of their interest while being self-beneficial. The Conference Navigation was evaluated at the E-Learn 2007 Conference which involved several parallel sessions and large number of papers.

## C. *Evaluation Metrics*

With reference to the descriptions of interactive and non-interactive recommender systems in [46], *SARVE* can be classified as an interactive recommender because user interaction data is obtained within the *SARVE* recommendation model. In recommender systems research, it is assumed that a recommendation is successful if and only if the recommended item/resource is beneficial and if and only if the item preference matches the target user's preferences. Thus we focused on the quality of recommendations [4] and adopted three commonly used classification metrics, namely: precision, recall and f-measure to evaluate our proposed algorithm.

Precision (*P*) metrics measures a recommender algorithm's ability to show only useful items, while it tries to minimize a combination of them with useless ones. Recall (*R*) metrics measures the coverage of useful items/resources the recommender algorithm/system can achieve. In other words, recall metrics measures the capacity of a recommender system/algorithm to obtain all useful items/resources present in the pool [44], [46].

Olmo and Gaudioso [46] summarized these facts using the confusion matrix (Table II) where *e* and *h* signify correct decisions (i.e. retrieve an item when it is relevant and do not when it is not). Additionally, *g* and *f* signify incorrect decisions (i.e. items should not be retrieved for recommendations). Equations (6) and (7) respectively depict the computations of precision and recall using variables *e, f* and *g*.

Classification metrics can be categorized into different recommendation outputs such as: (ii) *true positive* (*tp)*: an

TABLE II

CONFUSION MATRIX OF TWO CLASSES WHEN CONSIDERING THE RETRIEVAL OF DOCUMENTS/ITEMS

| Classes | Relevant Items | Irrelevant Items |
|---|---|---|
| Retrieved | *e (true positive)* | *f (false positive)* |
| Not Retrieved | *g (false negative)* | *h (true negative)* |

interesting item is recommended to the user, (ii) *true negative* (*tn)*: an uninteresting item is not recommended to the user, (iii) *false negative* (*fn*): an interesting item is not recommended to the user) and (iv) *false positive* (*fp)*: an uninteresting item is recommended to the user [44]-[46]. Therefore a more reliable recommender algorithm reduces the number of false negatives of users in order to achieve high values of recall and decrease false positives in order to obtain higher precision values.

Using (8), the f-measure ($F$) metric is used to simplify precision and recall into a single metric by blending their weights into absolute values.

$$P = \frac{e}{e+f} = \frac{good\ venues\ recommended}{all\ venue\ recommendations} \quad (6)$$

$$R = \frac{e}{e+g} = \frac{good\ venues\ recommended}{all\ good\ venues} \quad (7)$$

$$F = \frac{2 \times P \times R}{P+R} \quad (8)$$

*D. Evaluation Results*

To evaluate *SARVE*, we answer the following questions:

- What is the overall performance of *SARVE* in comparison to the other methods?
- What is the effect of cold-start and data sparsity in *SARVE*?

Furthermore, in (6) and (7), "*good venues recommended*" are classified as presentation sessions that corroborate similar tagged ratings and strong social ties between participants and presenters and as such fall within the social recommendation thresholds. Consequently, "*all venue recommendations*" and "*all good venues*" are relative in accordance to the different recommendation entity ranges in the dataset.

In terms of precision, both social context and social relation recommendations for *SARVE* were more precise and exact as measured by high Pearson correlation and social tie values. Referring to Fig. 7(a), at the highest value for Pearson correlation (1.0), *SARVE* achieved a higher precision (0.096) in comparison to that of B1 (0.075) and B2 (0.045). Similarly, in Fig. 7(b), at the highest value for social ties (0.8), *SARVE* attained a higher precision of 0.013 in comparison to that of B1 (0.0013) and B2 (0.0011). These scenarios indicate that *SARVE* displays more useful and exact items (presentation session venues) in comparison to B1 and B2.

Both social context recommendations and social relation recommendations for *SARVE* exhibited higher recall values and covered more useful items in accordance to the dataset. Referring to Fig. 8(a), at the highest value for Pearson correlation (1.0), *SARVE* attained a higher recall value of 0.810 in comparison to B1 (0.759) and B2 (0.698). Similarly, in Fig. 8(b), at the highest value for social ties (0.8), *SARVE* achieved a higher recall (0.809) in comparison to that of B1 (0.769) and B2 (0.728). In this case study, *SARVE* was able to execute a higher coverage of useful items (presentation session venues) within the pool in comparison to B1 and B2.

Consequently, according to Fig. 8, an upsurge in the values of Pearson correlation for the generation of social context recommendation and social ties for the generation of social relations recommendation will result in the increment of recall which will in effect increase *SARVE*'s ability to retrieve a higher coverage of useful items (presentation session venues) for participants. After computing the results of precision and recall metrics, we further computed their f-measure, these results as shown in Fig. 9. Fig. 9 corresponds to the precision and recall results obtained in Fig. 7 and Fig. 8. The results shown in Fig. 9 depict that *SARVE* outperformed B1 and B2 in terms of f-measure and this demonstrates its robustness and strength in terms of information retrieval in accordance to the dataset.

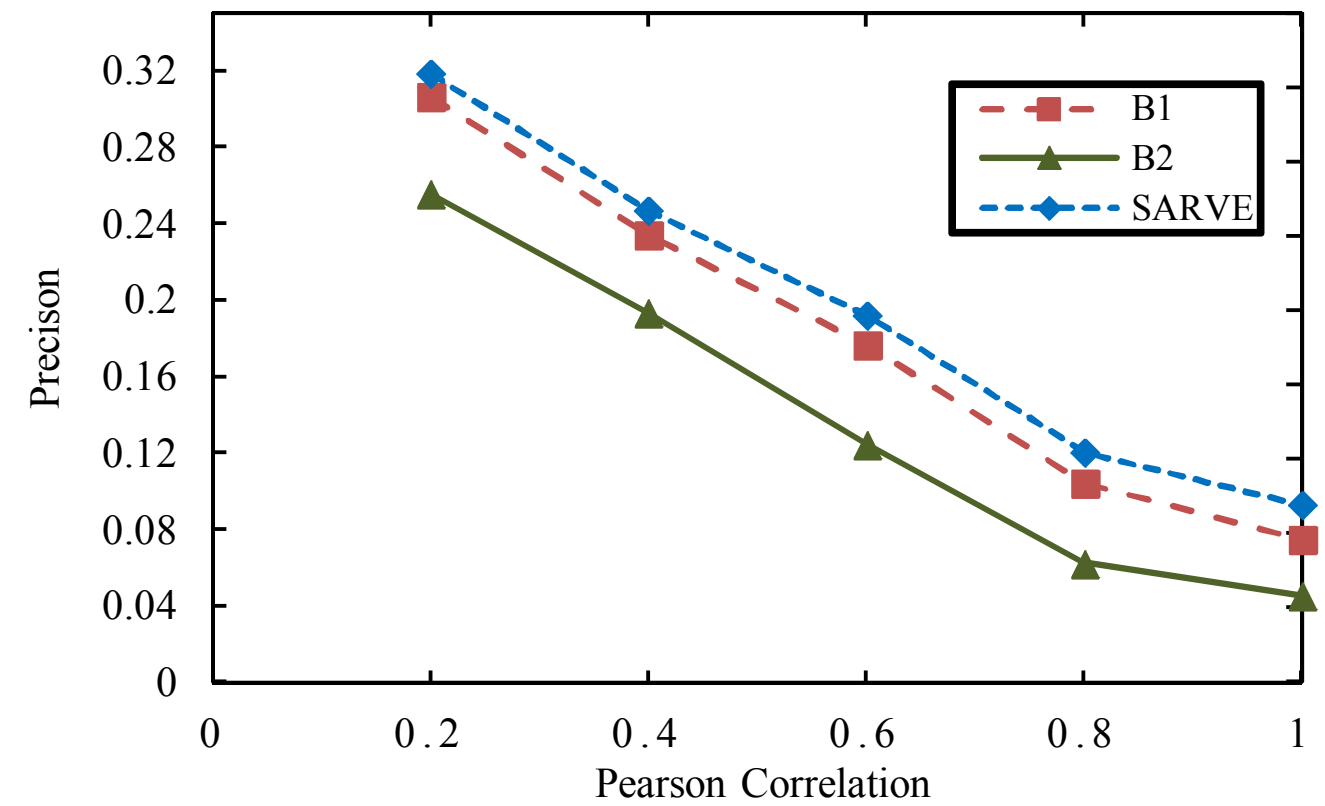

(a) Social context recommendation

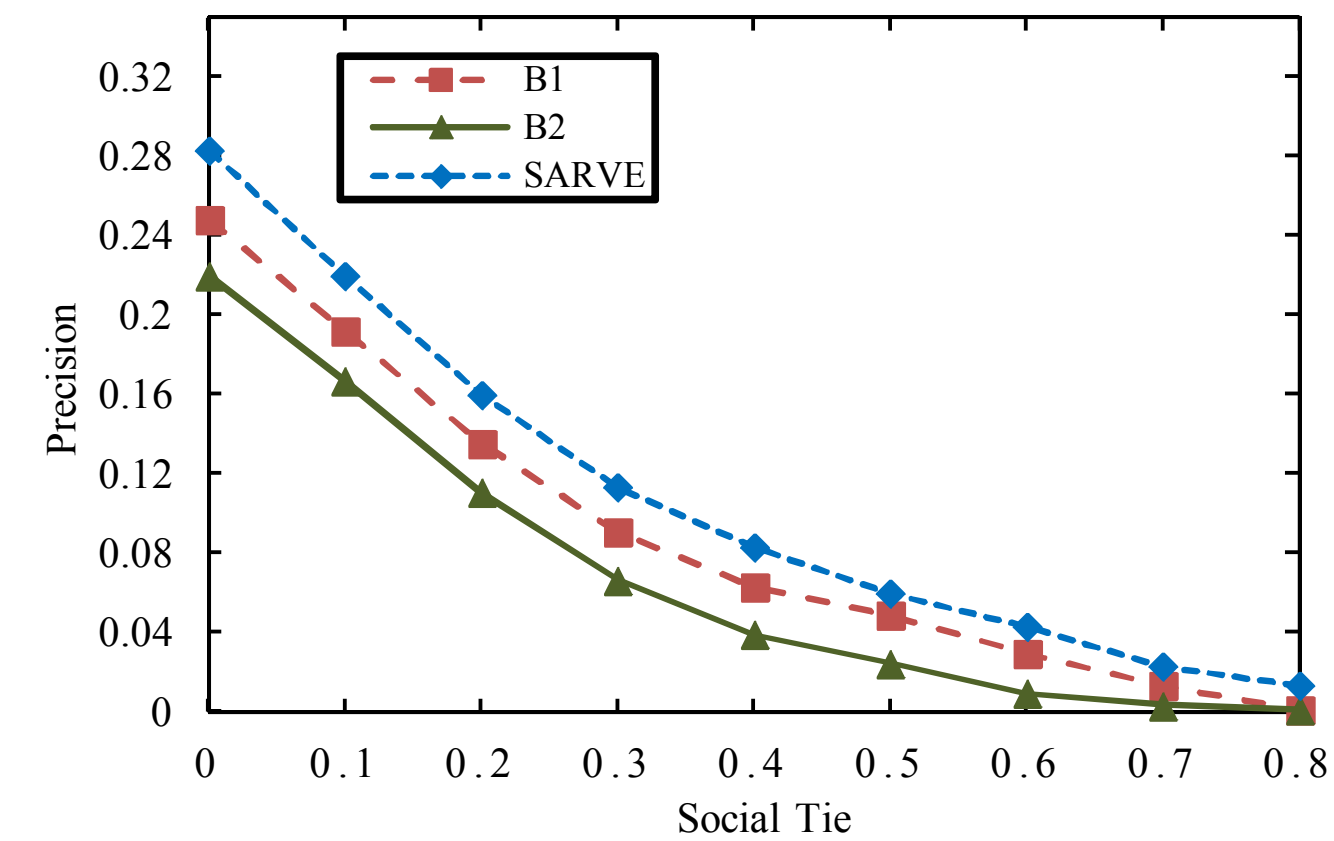

(b) Social relations recommendation

Fig. 7. Precision performance for ICWL 2012 dataset.

The *SARVE* described in this paper utilizes socially-aware recommendation through the integration of some social properties of conference participants. In comparison to B1 and B2, *SARVE* establishes a community detection approach for presentation session venues at the smart conference. Due to the effective utilization of contextual and social characteristic information pertaining to the smart conference environment, the algorithm outperforms both B1 and B2. B1 and B2 utilize Pearson correlation and B1 further utilizes social network analysis and link prediction, but the incorporation of the social properties illustrates the performance of the approach. Additional evidences appear in Tables III and IV.

By reinforcing user ratings and ensuring that conference participants are connected through a network of trust and social relationships, our method reduces problems of cold-start [47], [48] and data sparsity [49], [50]. Our *SARVE* approach generates two recommendations (social context and social relation) which are independent of each other due to differences in recommendation entities. Therefore in a scenario where a conference participant doesn't have common tagged patterns with a presenter, he/she may have strong social ties with the presenter and can be recommended a presentation session venue. Similarly, in another situation where the participant doesn't have strong social ties with a presenter, a social recommendation can still be generated for him/her based on similar tagged patterns with a presenter.

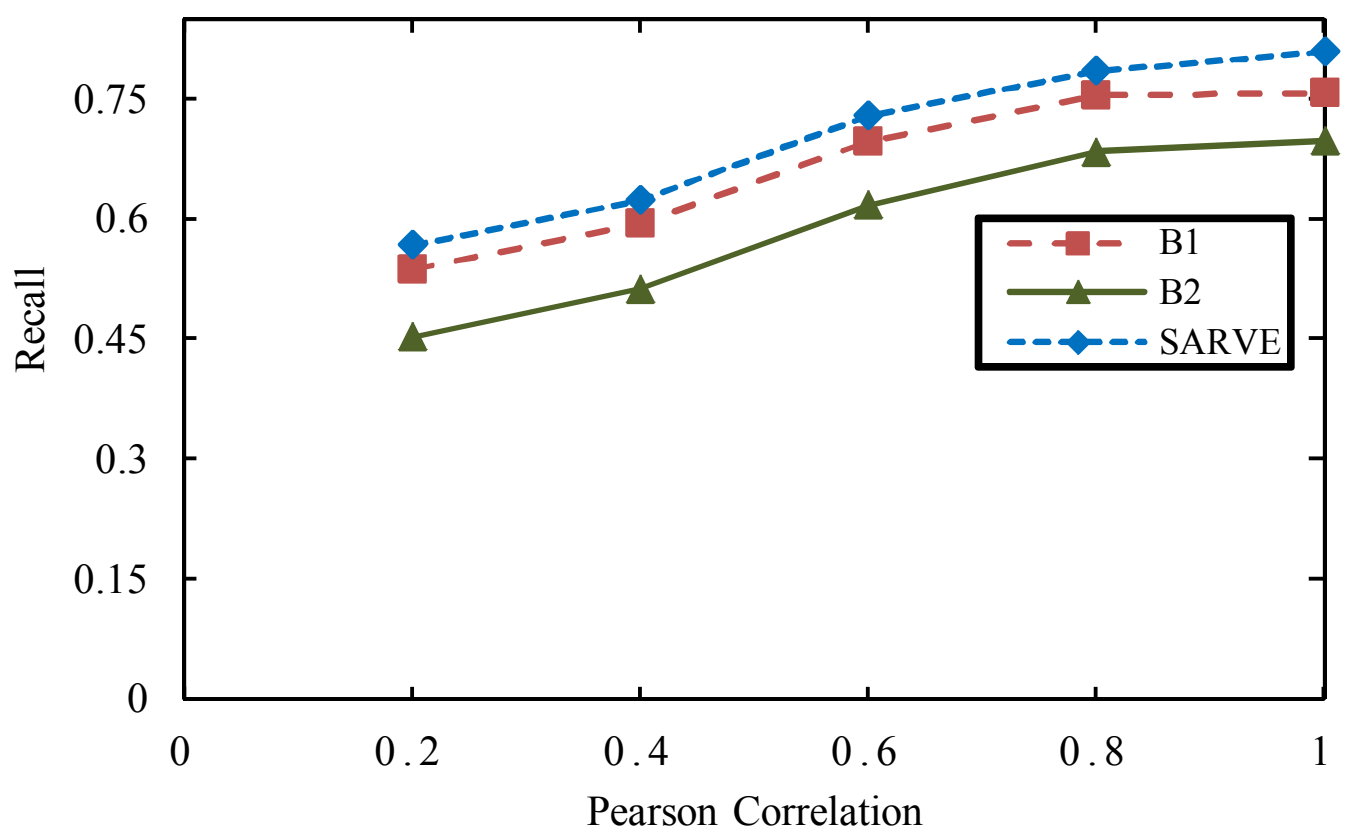

(a) Social context recommendation

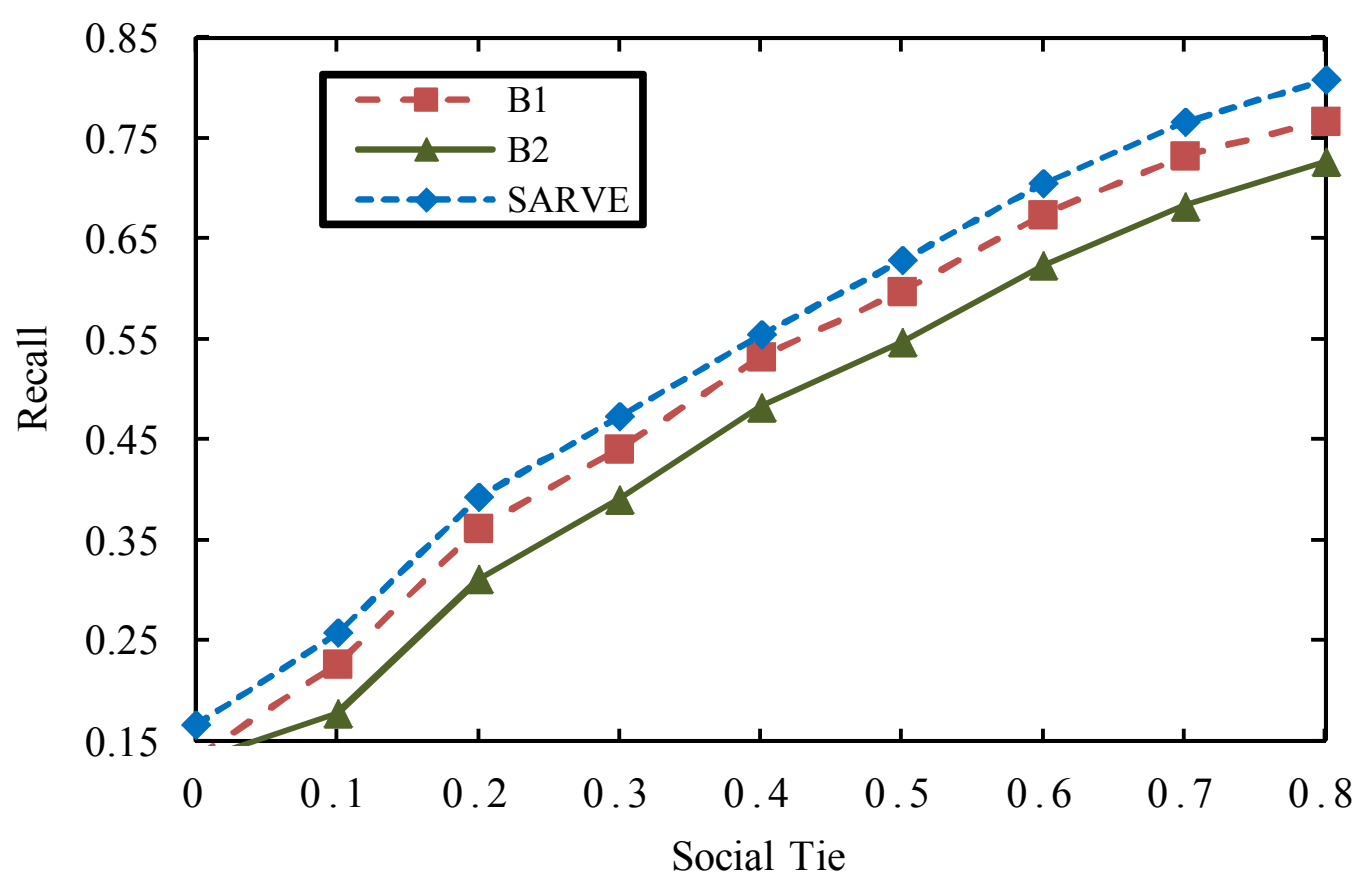

(b) Social relations recommendation

Fig. 8. Recall performance for ICWL 2012 dataset.

TABLE III
COMPARISON OF PROPOSED ALGORITHM IN TERMS OF PRECISION, RECALL AND F-MEASURE FOR SOCIAL CONTEXT RECOMMENDATION

| Algorithm | Highest Pearson | Precision | Recall | F-Measure |
|---|---|---|---|---|
| B1 | 1.0 | 0.075 | 0.759 | 0.137 |
| SARVE | 1.0 | 0.096 | 0.810 | 0.172 |
| B2 | 1.0 | 0.045 | 0.698 | 0.086 |

TABLE IV
COMPARISON OF PROPOSED ALGORITHM IN TERMS OF PRECISION, RECALL AND F-MEASURE FOR SOCIAL RELATIONS RECOMMENDATION

| Algorithm | Highest Social Tie | Precision | Recall | F-Measure |
|---|---|---|---|---|
| B1 | 0.8 | 0.0013 | 0.769 | 0.0026 |
| SARVE | 0.8 | 0.013 | 0.809 | 0.026 |
| B2 | 0.8 | 0.0011 | 0.728 | 0.0022 |

## VI. DISCUSSION

This paper presented a socially-aware recommendation approach that can be used to improve smart conference participation. We proposed an algorithm called *SARVE*, which recommends presentation session venues for participants at a smart conference. Using data consisting of context, social characteristics and research interests obtained through a relevant dataset, we were able to identify neighbors (participants who have similar interests and targets). We used this information as a guide to detect relevant communities pertaining to presentation session venues at the smart conference for the users (participants). Social ties and degree centrality were the social properties of users computed as part of the recommendation process. These measures were combined with dynamic explicit preferences and context of users in relation to presentation sessions at the conference in order to generate social recommendations for participants.

Results show that our approach is capable of providing useful social recommendations to conference participants and, for the example dataset, outperforms other state-of-the-art methods. Nevertheless, we observed that a limitation of *SARVE* may occur in cases where participants are recommended good presentation session venues through both strong social ties and high similarities of research interest (tagged) ratings. In such scenarios they have to decide which one is more suitable as they cannot be in two venues at the same time.

In the future, we would like to evaluate *SARVE* in more smart conferences to verify different impacts of recommender information on the quality of social recommendations gained through experimental threshold parameters. To achieve this target, location and proximity sensing instruments as well as computation of other social properties such as closeness and betweenness centrality must be explored to determine their possible availability at a smart conference venue. Additionally, using past and present social tie data, we would like to compute a more accurate prediction of social ties of conference participants and combine it with their personality traits. This will further improve accuracy in generating social recommendations for participants at smart conferences.

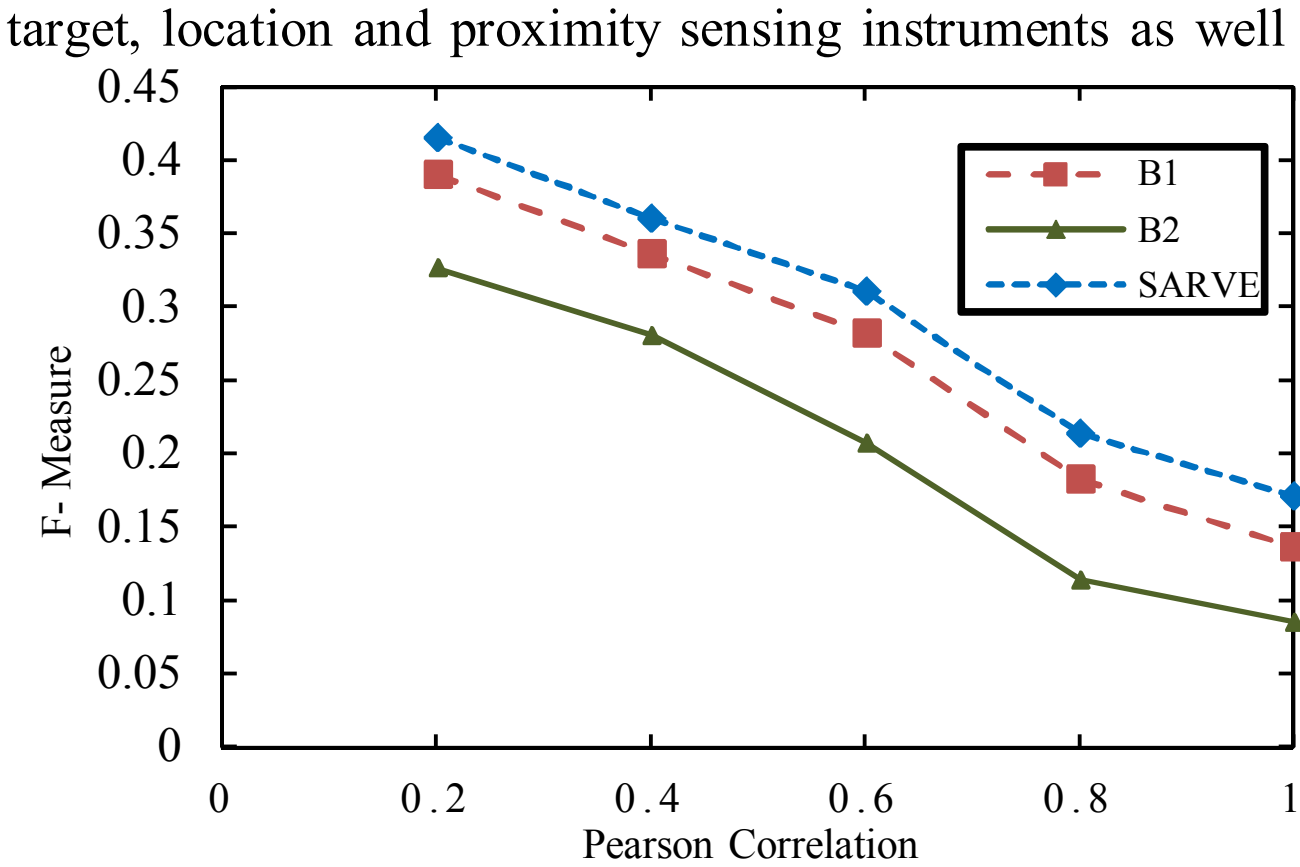

(a) Social context recommendation

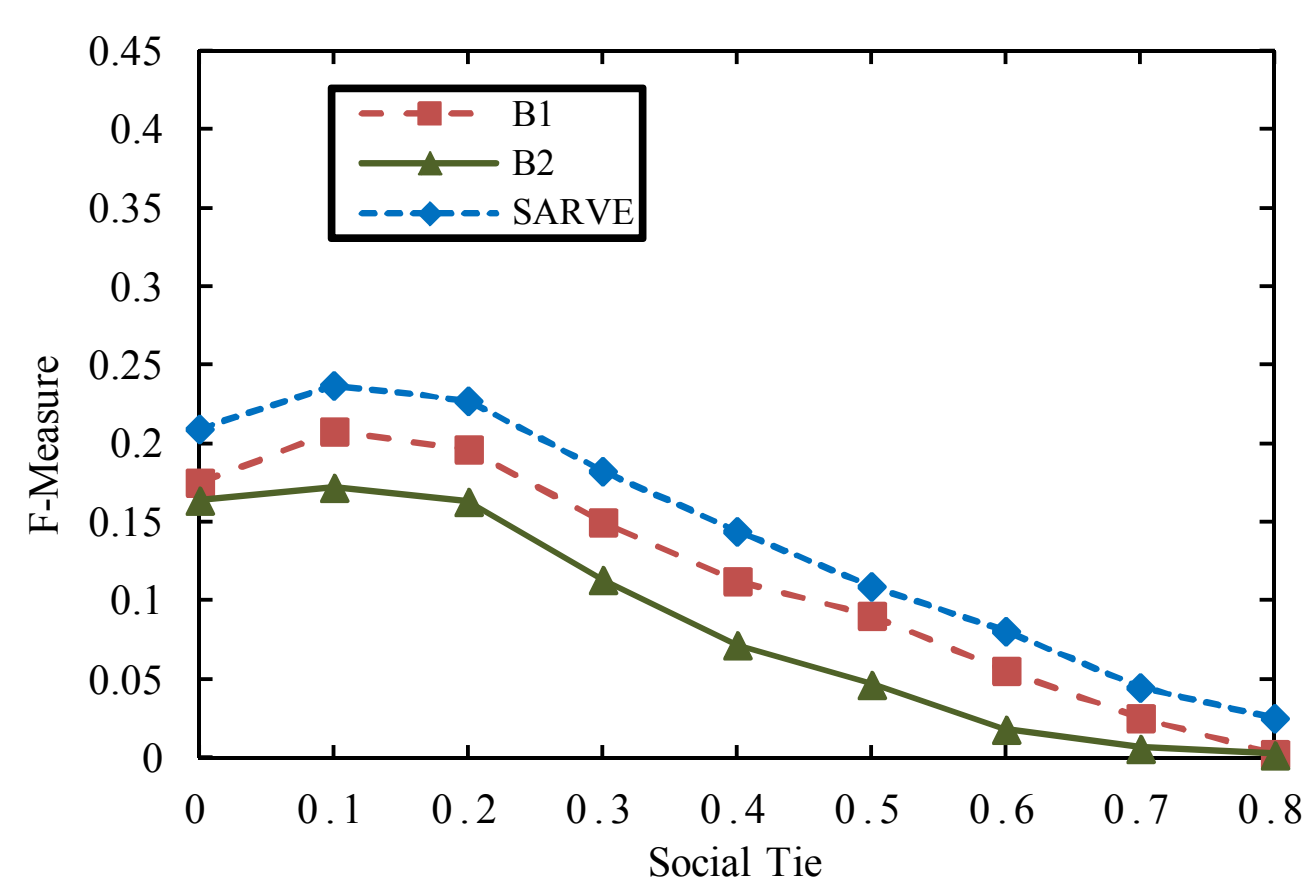

(b) Social relations recommendation

Fig. 9. F-Measure performance for ICWL 2012 dataset.

## Acknowledgement

The authors would like to thank Prof. E. Bass, the Editor in Chief and the reviewers for their constructive comments.